\documentclass[aps,preprint,onecolumn,superscriptaddress,preprintnumbers,12pt]{revtex4} 

\usepackage{amsfonts}
\usepackage{amsmath}
\usepackage{amssymb}
\usepackage{graphicx}%
\usepackage{bm}
\usepackage{subfig}
\usepackage{color}

\newcommand{\comments}[1]{}

\makeatletter
\renewcommand\@dotsep{10000}
\makeatother

\usepackage{mathrsfs}
\begin{document}


\title{The influence of transition metal solutes on dislocation core structure and values of Peierls stress and barrier in tungsten.}


\author{G.~D.~Samolyuk}
\affiliation { Materials Science and Technology Division, Oak Ridge National Laboratory, Oak Ridge, TN 37831, USA } 

\author{Y.~N.~Osetsky}
\affiliation { Materials Science and Technology Division, Oak Ridge National Laboratory, Oak Ridge, TN 37831, USA } 

\author{R.~E.~Stoller}
\affiliation { Materials Science and Technology Division, Oak Ridge National Laboratory, Oak Ridge, TN 37831, USA } 

\date{4 February 2009}


\begin{abstract}

Several transition metals were examined to evaluate their potential for improving the ductility of tungsten. The dislocation core structure and Peierls stress and barrier of $1/2\langle111\rangle$ screw dislocations in binary tungsten-transition metal alloys (W$_{1-x}$TM$_{x}$) were investigated using first principles electronic structure calculations. The periodic quadrupole approach was applied to model the structure of $1/2\langle111\rangle$ dislocation. Alloying with transition metals was modeled using the virtual crystal approximation and the applicability of this approach was assessed by calculating the equilibrium lattice parameter and elastic constants of the tungsten alloys. Reasonable agreement was obtained with experimental data and with results obtained from the conventional supercell approach. Increasing the concentration of a transition metal from the VIIIA group, i.e. the elements in columns headed by Fe, Co and Ni, leads to reduction of the $C^\prime$ elastic constant and increase of elastic anisotropy A=$C_{44}/C^\prime$. Alloying W with a group VIIIA transition metal changes the structure of the dislocation core from symmetric to asymmetric, similar to results obtained for W$_{1-x}$Re$_{x}$ alloys in the earlier work of Romaner {\it et al} (Phys. Rev. Lett. 104, 195503 (2010))\comments{\cite{WRECORE}}. In addition to a change in the core symmetry, the values of the Peierls stress and barrier are reduced. The latter effect could lead to increased ductility in a tungsten-based alloy\comments{\cite{WRECORE}}. Our results demonstrate that alloying with any of the transition metals from the VIIIA group should have similar effect as alloying with Re.
\end{abstract}


\maketitle



\section{Introduction}

\comments{Tungsten has the highest melting temperature of any metal and is the prime candidate for use in the divertor of future fusion reactors~\cite{FR}.}
{\color{black} Tungsten is the prime candidate for use in the divertor of future fusion reactors because of its high melting temperature and resistance to sputtering~\cite{FR, SPUTTERING}.} However, its lack of ductility is an impediment to its use. The low-temperature brittleness is a common problem for all metals from the VIA group, such as chromium, molybdenum and tungsten~\cite{KLOPP}. The ductility of these metals can be improved by alloying with rhenium, leading to the so-called "Re effect"~\cite{REEFFECT, KLOPP, Savitskii}. However, Re, is a very rare and expensive element. It is therefore desirable to find alternate elements which provide a similar increase in ductility at lower cost. An experimental investigation of the range of candidates in the periodic table would be rather expensive, but computational materials science methods based on accurate first-principles calculations provide a very promising way to narrow the range of possible candidates.  

Several possible mechanisms for the Re-effect have been discussed~\cite{GILBERT,KLOPP,Raffo,Luo,GORN1,Trefilov,Kurdyumova}, and the following two mechanisms are selected as most promising in application to monocrystals, i) solid solution softening in which an impurity improves mobility of 1/2(111) screw dislocation and ii) enhancement of cross-slip in which an impurity modifies the dislocation core structure making it easier to cross-slip and increasing the number of possible slip planes. Both of these are related to the effect of impurities on the dislocation core and may be amenable to investigation by first-principles computational methods even though such calculations cannot be used to directly estimate the mechanical properties. Therefore, it should be possible to define one or more calculable figures of merit that are related to a material’s elastic properties and potentially to ductility. Among the potential figures of merit are the material’s individual elastic constants, Poisson’s ratio~\cite{GAO}, and Peierls stress; and there is strong evidence that these parameters are also directly related to the electronic structure of particular impurities.

\comments{and it was suggested~\cite{GORN1} that two groups of mechanisms could be distinguished as influencing the mechanical properties of specific single-crystalline and polycrystalline materials. One group deals with the effect of Re on the twinning and mobility of individual dislocations and the other addresses the effect of Re on dislocation mobility due to trapping of interstitial atoms and the suppression of impurity segregation on the dislocation and grain boundaries, a "scavenging" mechanism~\cite{scavenging,Medvedeva}. However, scavenging mechanisms do not explain all the experimental observations and therefore in this paper we concentrate our efforts on the first group.}

The experimentally observed strong correlation between the number of valence electrons on the solute atom and the degree of softening~\cite{Hiraoka_Mo,Klopp2,Stephens1975265} points to the importance of electronic factors in solution softening~\cite{Medvedeva0, Medvedeva}.  First principles calculation~\cite{WRECORE, WRECORE_2}  indicate that the ductilizing effect of Re could originate from "direct" improvement of the mobility of $1/2\langle111\rangle$ screw dislocations~\cite{Raffo,Trinkle,Medvedeva0,Medvedeva}, either by decreasing the Peierls stress, or enhancing cross-slip by changing the slip plane from $\{110\}$ to $\{112\}$ which would increase the number of available slip planes from 6 to 12~\cite{GILBERT,Garfinkle}. First principles density functional (DFT) calculations demonstrated that alloying with Re in a W$_{1-x}$Re$_x$ alloy leads to a transition of the $1/2\langle111\rangle$ screw dislocation structure from the symmetric core to asymmetric core, and to a reduction in Peierls stress~\cite{WRECORE}. {\color{black} Closely related results were obtained from first-principles calculation for Mo alloys with 5$d$ transition metals~\cite{Trinkle}. The authors~\cite{Trinkle} placed solutes in a row along the $1/2\langle111\rangle$ dislocation core and calculated the change of stiffness associated with moving the row along the $\langle111\rangle$ direction. According to the results, solutes having fewer $d$ electrons (Hf and Ta) increase the stiffness, which authors infer strengthens the Mo alloy, whereas those having more $d$ electrons (Re, Os and Ir) decrease stiffness, leading to softening.} 

The goal of the present investigation is to study possible substitutes for Re in alloys that will result in a similar ductilizing effect as Re. The properties of W-transition metal (TM) alloys, were modeled using the virtual crystal approximation~\cite{Faulkner19821} (VCA). {\color{black} The applicability of this approach to modeling elastic properties, structural stability and phonon properties of a W-Re alloy has been demonstrated~\cite{GORN1,DYN_INSTAB,PHASE_DIAGR}. The approach we use can be described as follows. First, a dislocation is placed in the effective media representing the W-TM alloy. \comments{The properties of the dislocation in this media are calculated exactly with the actual W-TM alloy is described by a periodic lattice with lattice parameter chosen to satisfy zero pressure condition and real pseudopotentials of impurity and host atoms are substituted by averaged pseudopotentials with averaged number of electrons.} The properties of the dislocation in this medium are calculated exactly but for a W-TM alloy described by a pseudopotentail which is the weights average of the actual pseudopotentials of the impurity and host atoms and the averaged number of electrons. A periodic lattice is used with the lattice parameter chosen by obtain zero pressure condition. This approach corresponds to a zero order contribution to the electronic system energy expansion with respect to the difference between real atomic potentials and the virtual atom. Such an approach allows separating the so-called "band structure effects", in this particular case filling of $d$-states, from the effect of local modification of the lattice due to W substitution. The latter is not possible in a direct super cell calculation. It was demonstrated~\cite{WRECORE} that dislocation properties are sensitive to modifications of this effective media and alloying with Re leads to a sizable reduction of Peierls stress and barrier. Within this approach, a comparison of elastic constants calculated using the VCA and super-cell methods provides a verification of the VCA accuracy as we demonstrate below.}

We show that alloying with transition metals with a higher number of $d$ electrons (VIIIA group) reduces the Peierls stress and barrier for a $1/2\langle111\rangle$ screw dislocation. Similar to W-Re alloys, the dislocation core symmetry is reduced by alloying with TM from group VIIIA. It is demonstrated that the scale of Peierls barrier reductions are similar for all W$_{1-x}$TM$_x$ alloys with the same ratio of electrons per atom ($e/a$). {\color{black} Although, the VCA approach} describes an influence of band-structure on the properties of screw dislocations, it does not describe the discrete nature of alloy structure. However, this approach permits an assessment that can significantly reduce the range of possible solute candidates suitable for Re substitution. This paper is organized as follows. In section \ref{sec:approaches} we briefly review computational details used in the calculations. Section \ref{sec:el_str} describes the electronic structure of tungsten and its alloys with transition metals from IVA, VA and VIIA groups. This is an almost complete list of possible binary solid solutions~\cite{TUNGSTEN}. The results of modifications to the elastic constants with alloying are discussed in section \ref{sec:elastic}. Section \ref{sec:dislocation} presents our prediction of core structure, Peierls barrier and Peierls stress for W$_{1-x}$TM$_x$ alloys.  Finally, the conclusions are given in section \ref{sec:conclusions}.

\section{Computational approaches}
\label{sec:approaches}

The electronic structure within the generalized gradient approximation (GGA) of density functional theory (DFT) was calculated using the {\it QUANTUM ESPRESSO} (QE) package~\cite{PWSCF}. The calculation was done using a plane-wave basis set and ultrasoft pseudo-potentials optimized in the RRKJ scheme~\cite{RRKJ}. We used the Perdew-Wang~\cite{PW91} exchange-correlation functional. The Brillouin zone (BZ) summations were carried out over a $24\times24\times24$ BZ grid for the system with one unit cell and $16\times16\times16$ grid for the supercell containing $2\times2\times2$ unit cells, with electronic smearing with a width of 0.02 Ry applied according to the Methfessel-Paxton method. The plane wave energy cut off of 42 Ry allows reaching an accuracy of 0.2 mRy/atom. As a realization of VCA for the pseudo potential method, we used the scheme proposed by Ramer and Rappe~\cite{VCA}. The elastic constants were calculated from the total energies obtained for the set of unit cell deformations~\cite{C_IJ_CALC}.

\begin{figure}
\includegraphics[angle=0,width=7.0cm]{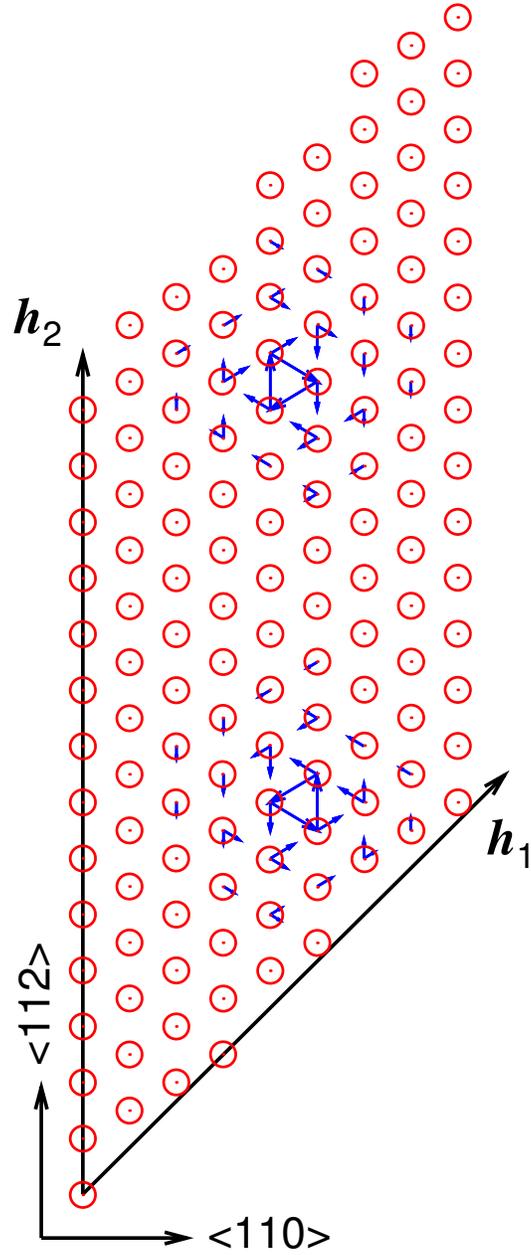}
\caption{The [111] projection of dislocation dipole unit cell, where atoms are shown by circles and arrows corresponds to differential displacements.}
\label{SC}
\end{figure}

We use a periodic quadrupolar arrangement for a $1/2\langle111\rangle$ screw dislocation~\cite{QUADRUPOL_Si} in the cell with basis vectors $\vec b_1 = 9 \vec u_1$, $\vec b_2 = 5 \vec u_2$ and  $\vec b_3 = \vec u_3$, where $\vec u_1=[\bar{1}10]$, $\vec u_2=[\bar{1}\bar{1}2]$ and $\vec u_3 = 1/2 [111]$. An appropriate choice of lattice vectors~\cite{QUADRUPOL_Si} reduces the quadrupole cell to half the size, and we therefore use a cell with basis vectors equal $\vec h_1 = (\vec u_1 + \vec u_2 + \vec u_3)/2$, $\vec h_2 = \vec b_2$ and  $\vec h_3 = \vec b_3$ and 135 atoms. This unit cell contains only two dislocations with opposite Burgers vectors (Fig.~\ref{SC}). It was demonstrated earlier that this cell size is large enough to reproduce such characteristics of the dislocations as Peierls stress and barrier~\cite{PEIERLS_SMALL_CELL, Arias_Mo_Ta, Arias_Ta, Frede_Jacobsen, Bulatov_Yip, Willaime, Od} reasonably well {\color{black} if the elastic interaction correction is included. We are interested in how the barriers change with solute concentration and therefore the above correction is not included since it's the same for all concentrations.} For the dislocation calculation, the BZ summation was carried out over a $1\times2\times8$ BZ grid~\cite{WRECORE} and the initial structure was relaxed until the forces were smaller than 0.0005 Ry/\AA.

\section{Results}

\subsection{Electronic structure}
\label{sec:el_str}

\begin{figure}
\includegraphics[angle=270,width=16.0cm]{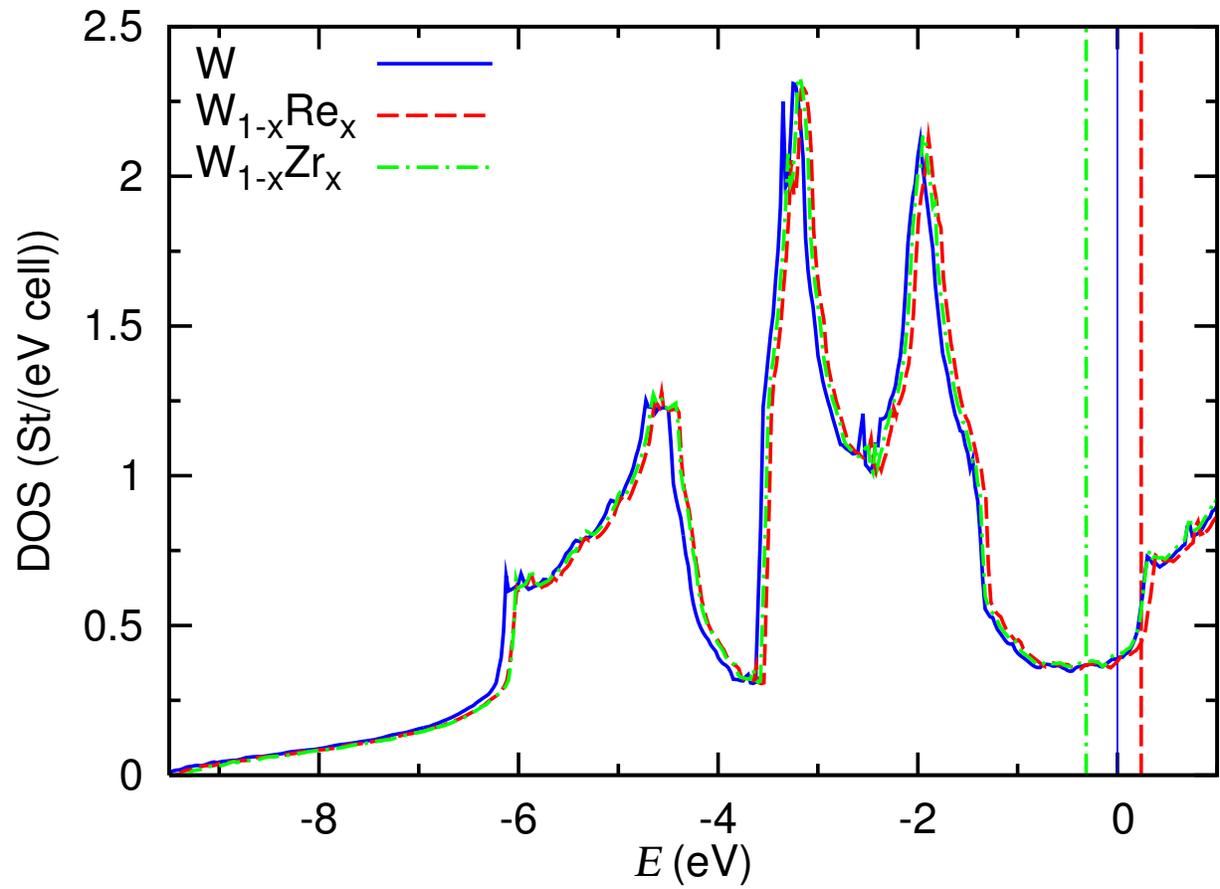}
\caption{DOS for pure W, blue solid line, W$_{0.9375}$Re$_{0.0625}$, red dashed line, and W$_{0.9375}$Zr$_{0.0625}$, green dash-dotted line, calculated using VCA.}
\label{DOS_VCA}
\end{figure}

The electronic density of states (DOS) calculated using the VCA are presented in Figure~\ref{DOS_VCA} (colored online) for pure tungsten (blue solid line) and two tungsten alloys, one with 6.25 \% Re (red dashed line) and the second one with 6.25 \% Zr (green dash-dot line). Zero energy on this plot corresponds to occupation of electronic bands by six electrons. Thus, for pure tungsten, zero energy corresponds to the Fermi level. The Fermi energy is placed in the pseudo gap between bonding and antibonding $d$-states. As can be seen from Fig.~\ref{DOS_VCA}, alloying with Re or Zr doesn't produce sizable changes in the DOS, at least in VCA approach. The substitution of tungsten with 6.25 \% Re leads to an upwards shift of the Fermi energy, shown by a red vertical line. The area under DOS between zero energy and the W$_{0.9375}$Re$_{0.0625}$ Fermi energy corresponds to 0.0625 additional electrons, i.e., occupation of each virtual atom by an additional 0.0625 electrons. Correspondingly, alloying with atoms that have a lower number of electrons, such as Zr, leads to a downward shift of the Fermi energy, shown by vertical green line, and each virtual atom is occupied by 0.0625 fewer electrons.




Such a modification of electronic structure in tungsten alloys can be described using the rigid band approximation~\cite{Faulkner19821}. This approach assumes that alloying does not change the DOS but shifts the Fermi level, so the occupation of each virtual atom corresponds to the average number of electrons per atom $(e/a)$ in the alloy. Thus, within this approximation, all alloys with the same $(e/a)$ value have the same properties. The concentration of different alloying elements which have the same $(e/a)$ value can be calculated through the simple expression
\begin{equation}
x = \frac{(e/a)}{Z_{TM}-Z_W}
\label{x_vca}
\end{equation}
, where $Z_{TM}$ is the number of valence electrons of the alloying atom and $Z_W$  is the number for tungsten. Although this approach has limited accuracy, it can be very useful for qualitatively  estimating the change in such properties as elastic constants or Peierls barrier. It should be mentioned that alloying of tungsten with TM from VIIA and VIIIA groups, within their solubility limits, fills bonding $d$ states with additional electrons and increases strength of the bonds. {\color{black} It reflects in the increase of cohesive energy and reduction of Wigner-Seitz radius for 4$d$ TM with filling of bonding $d$ states and reduction of cohesive energy and increase of Wigner-Seitz radius with filling of antibonding states~\cite{Moruzzi}}.

\begin{figure}
\includegraphics[angle=270,width=16.0cm]{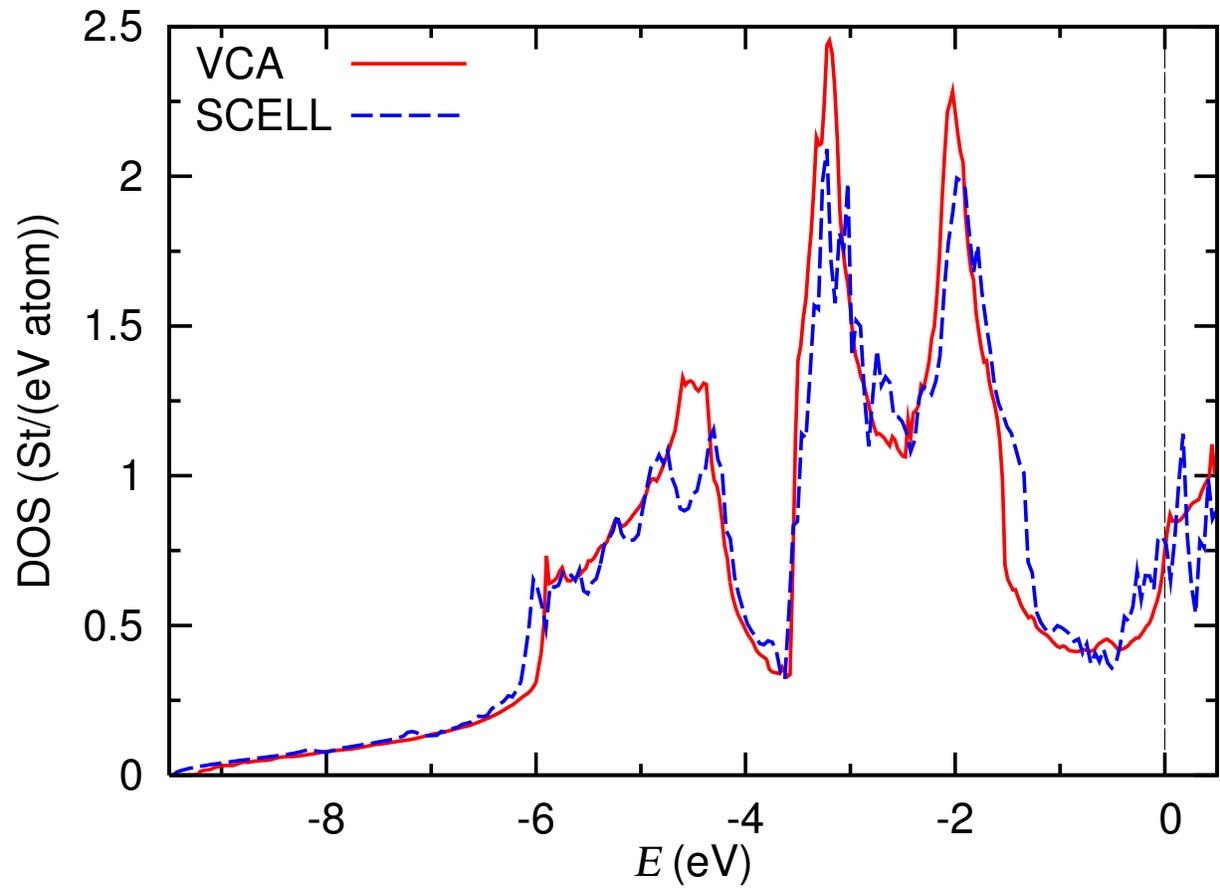}
\caption{W$_{0.9375}$Fe$_{0.0625}$ DOS per atom calculated using a $2\times2\times2$ supercell model, blue dashed line, and VCA, red solid line. The Fermi energy corresponds to zero.}
\label{DOS_SCELL}
\end{figure}

The VCA approach could be quite inaccurate for describing the properties of disordered substitutional metallic alloys, especially for transition metals~\cite{Faulkner19821}  and therefore the accuracy of this approach should be analyzed in each particular case. In order to verify the accuracy of VCA we compared its results with those obtained using a supercell approach. The supercell contained $2\times2\times2$ cubic unit cells with one tungsten atom substituted by a solute atom. Thus, the solute concentration corresponds to 6.25 \%. In Figure~\ref{DOS_SCELL}, the DOS calculated using the supercell (blue line) and VCA models (red line) are presented for W$_{15}$Fe.
The DOS calculated using the VCA approach is very close to the supercell result. \comments{It should be mentioned that thin structure of supercell calculated DOS around Fermi level is a drawback of the method and it's more correct to compare VCA result with averaged supercell DOS.} Comparisons of elastic properties calculated using the two approaches will be discussed below.

\subsection{Elastic constants}
\label{sec:elastic}

The modification of elastic properties in W-Re alloys has been widely investigated both experimentally~\cite{C_IJ_WRE} and theoretically~\cite{DYN_INSTAB,PHASE_DIAGR}. Here we expand these investigations to a wide set of tungsten-based alloys. The results of equilibrium lattice constant, $a$ , bulk modulus, $B$, and elastic constants, $C_{11}$, $C_{12}$, $C_{44}$ and $C^\prime=(C_{11}-C_{12})/2$, calculated by both VCA and supercell methods, are presented in  Table \ref{elst_cnst_vca_vs_sc}. For the elastic constants, the largest difference between calculated and experimental values was obtained for $C_{44}$ and is about 13 \% for the VCA calculations and slightly larger for $2\times2\times2$ super cell calculations (SC). 
{\color{black}
The agreement between VCA and SC results W alloys with Zr, Ta, and Re is very good and differences for elastic constants do not exceed 2 \%. Hoever this difference increases for alloys with solute atoms with a larger number of valence electron. For example, for the Fe-family (Fe, Ru and Os) the difference is about 5 \%, while for the Co- (Co, Rh and Ir) and for Ni-families (Ni, Pd and Pt) the difference increases to 12-15 \%. Since we investigate the reduction of Peierls stress and barrier for alloys in which the number of valence electrons per atom $(e/a)$ corresponds to 10 \% of Re or smaller the elastic properties of the alloys should be described with an accuracy similar to Re of 5 \% or less.
As discussed above, alloying of tungsten with transition metals with a higher number of $d$ electrons, metals from the VIIA and VIIIA groups, fills bonding $d$ states. According to the general results obtained for transition metals~\cite{Moruzzi,Faulkner19821}, this filling leads to a decrease of the lattice parameter. The calculated results for W$_{1-x}$Re$_x$ alloys reproduce this general tendency (see Table \ref{elst_cnst}) and agrees with experiment~\cite{C_IJ_WRE}. The same tendency was obtained in tungsten alloyed with other TM except Pd and Pt. For these two elements, the lattice parameter calculated in SC slightly increases.  Another general trend for all tungsten alloys with solute atoms having a higher number of $d$ electrons is reduction of the $C^\prime$ elastic constant. The calculated results in Table~\ref{elst_cnst} are also in good agreement with experimental values for the W-Re alloy~\cite{C_IJ_WRE} and reflect a reduction in the stability of bcc structure. At concentrations higher than x=0.25\% the bcc W$_{1-x}$Re$_x$ transforms to the $\sigma$-phase~\cite{BINARY_PHDIAG}. Similar to previous results~\cite{DYN_INSTAB}, the $C^\prime$ modulus decreases with Re concentration until it becomes negative at 85 \% Re. This change in the sign of $C^\prime$ corresponds to the dynamic loss of stability of the bcc structure. The $C^\prime$ elastic constant corresponds to the long-wave transversal phonon branch in $[\xi\xi0]$ directions (T $[1\bar{1}0][\xi\xi]$) and softening of this phonon mode provides a transition path from the bcc to dhcp structure (see discussion in \cite{DYN_INSTAB}). 
}

\begin{figure}
\includegraphics[angle=270,width=17.0cm]{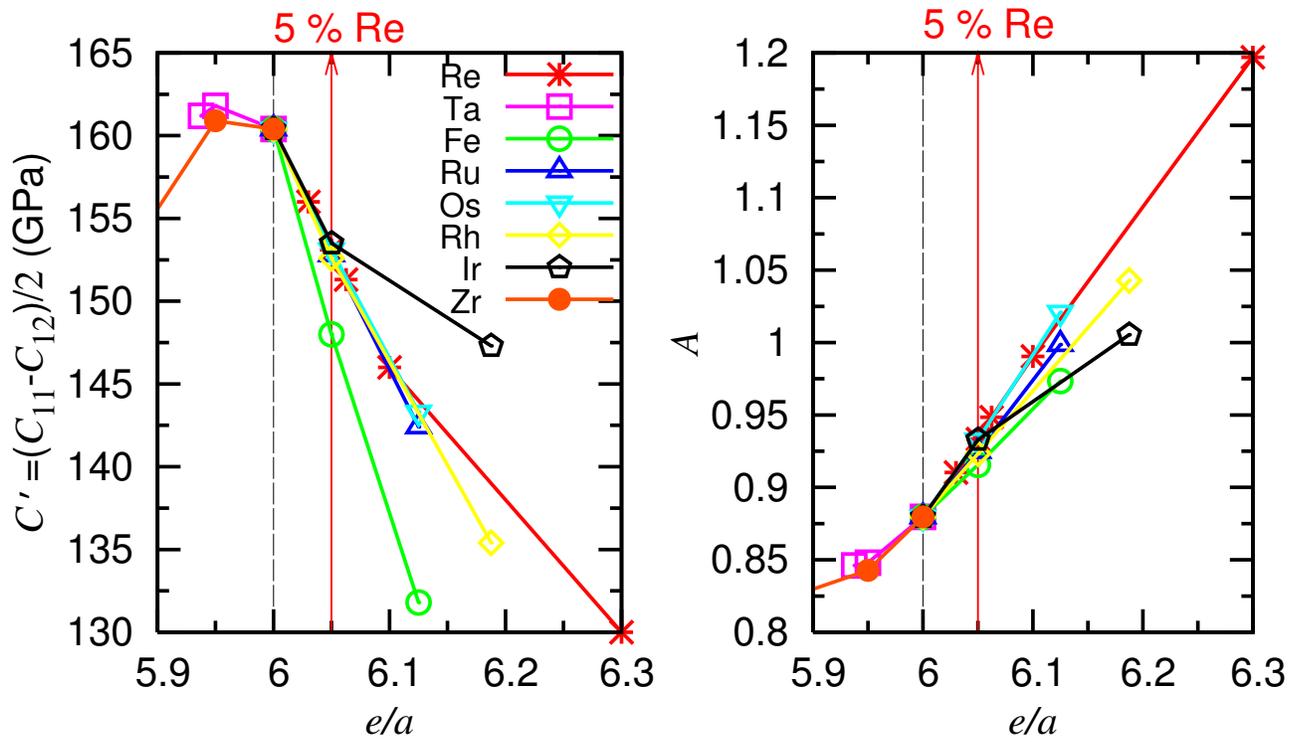}
\caption{$C^\prime$ and $A$ as a function of number of electrons per atom $e/a$ calculated for W$_{1-x}$TM$_{x}$ using VCA.}
\label{cp_a}
\end{figure}

According to the rigid band model discussed in the previous section, the parameter which determines the elastic properties of W$_{1-x}$TM$_x$ alloys is the number of electrons per atom $e/a=(1-x)Z_W+xZ_{TM}$ (see explanation of Eq.\ref{x_vca}). In Fig.~\ref{cp_a} the elastic constant $C^\prime$ and elastic anisotropy $A=C_{44}/C^\prime$ are presented as a function of $e/a$. The $C^\prime$ and $A$ values are very similar for all TM from group VIIA and VIIIA except for Fe and Ir. However, even for these two elements the deviation from the average values is around 7 \%.
Additional evidence for the importance of the $e/a$ parameter follows from the correlation between the width of the stable bcc solid solution region of W$_{1-x}$TM$_x$ and the number of valence electrons for TM elements from groups VIIA and VIIIA~\cite{BINARY_PHDIAG}. Alloys with solute atoms having a higher number of valence electrons have a narrower solid solution region.

\subsection{Dislocation core structure, Peierls stress and Peierls barrier.}
\label{sec:dislocation}

\comments{In bcc structures $1/2\langle111\rangle$ screw dislocations have two nonequivalent core configurations, that are referred to as "easy" and "hard"~\cite{Vitek, Xu_Moriarti_Mo, Arias_Mo_Ta} - or stable and metastable, respectively. These two configurations can be obtained for a given Burgers vector $\vec b$ by placing the core origin in specific positions.}

{\color{black} In bcc structures $1/2\langle111\rangle$ screw dislocations have three nonequivalent core configurations, that are referred to as "easy", "hard" and "split"~\cite{Vitek, Xu_Moriarti_Mo, Arias_Mo_Ta, TAKEUCHI_Fe} - stable and two metastable, respectively. These configurations can be obtained for a given Burgers vector $\vec b$ by placing the core origin in specific positions. Recently, it was demonstrated that in bcc Fe the lowest energy is the symmetric "easy" core configuration, then symmetric "hard" core and the highest energy is the "split" core configuration~\cite{Itakura_Fe}. {\color{black}In a contrast to Fe, for the case of W we determined that the symmetric "hard" core configuration decays into the "split" core}. \comments{\color{black}In a contrast to Fe, in the case of W we determined that the symmetric "hard" core has higher energy then the "split" core configuration.} Thus, in the present publication we assume that the dislocation migrates between the "easy" and the "split" core configurations.}

\begin{figure}
\includegraphics[angle=270, width=16.2cm]{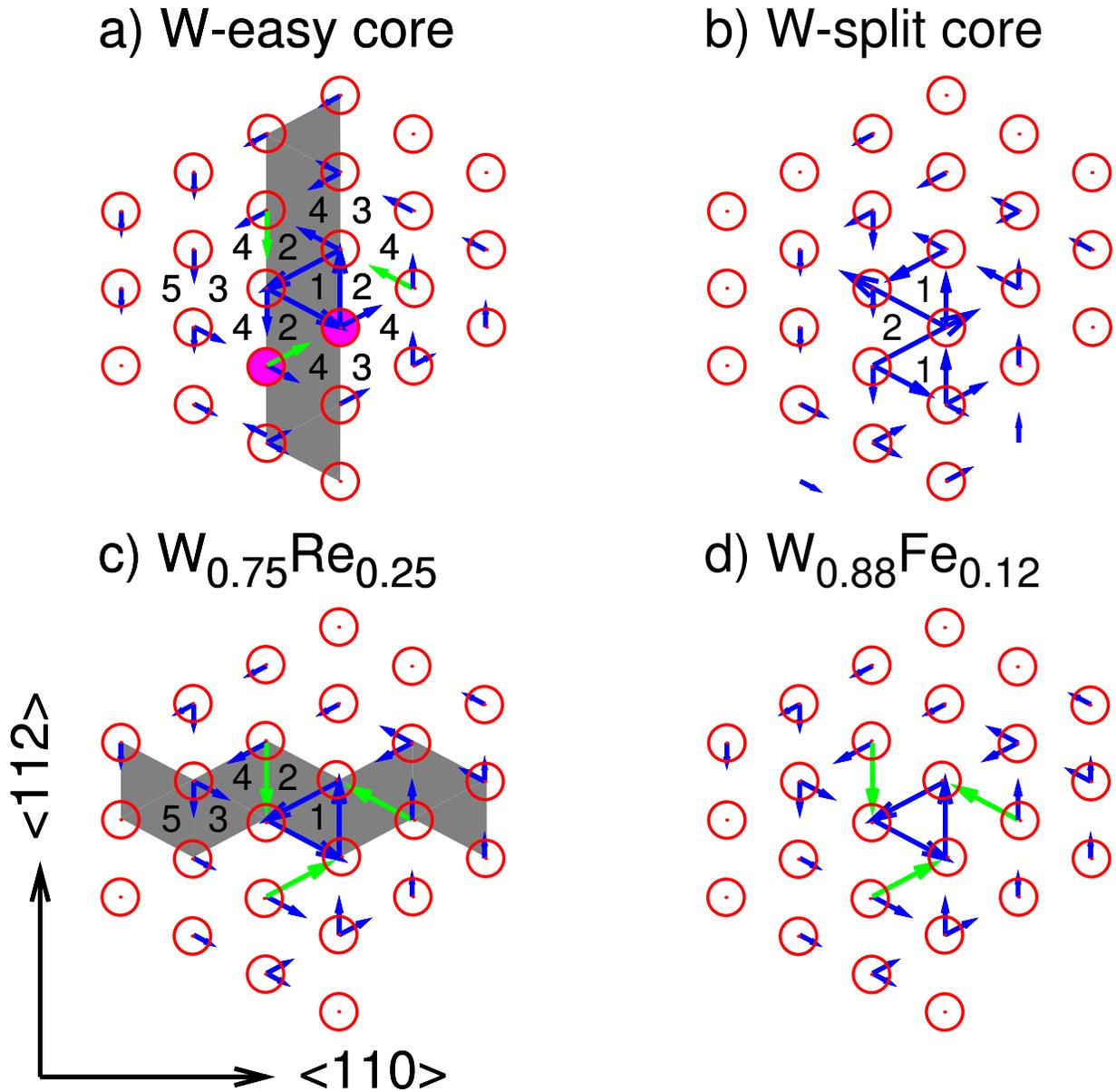}
\caption{\color{black} The DD map of "easy" core dislocation in W$_{1-x}$TM$_x$ calculated using the VCA approach. Some of arrows are highlighted in green to illustrate a change of core symmetry. The sites 1, 4 and 5 correspond to the stable "easy" core dislocation center, while sites 2 and 3 indicate the metastable "split" core center. The $z$-coordinate of atoms in columns shown by filled circles are fixed during calculation of Peierls barrier within the reaction coordinate method. The symmetric and asymmetric core glide paths are shown by gray bands.}
\label{fig:CORE_STR}
\end{figure}

Following, Xu and Moriarty~\cite{Xu_Moriarty} we marked these sites as 1, 4 or 5 for "easy" cores and site 2 or 3 for "split" cores in Figure  \ref{fig:CORE_STR}. The calculated structure of the "easy"  core configuration for a $1/2\langle111\rangle$ dislocation is shown in Fig. \ref{fig:CORE_STR} for pure W, W$_{0.75}$Re$_{0.25}$ and W$_{0.88}$Fe$_{0.12}$, where the concentrations of transition metals were chosen to give the same $(e/a)$ value. The circles in Fig. \ref{fig:CORE_STR} represent W atoms looking in the $\langle111\rangle$ direction and the dislocation structure is illustrated by differential displacement (DD) maps~\cite{Vitek}. On a DD map the displacement of atoms in the [111] direction relative to a neighbor is plotted as an arrow connecting the atom with its neighbors. The length of an arrow is normalized so that arrows connecting two atoms corresponds to displacement by $b/3$. Thus, summing arrows in the circuit around a dislocation gives the Burgers vector. For tungsten alloys {\color{black} with VIIIA group} TM having the same $(e/a)$ value, the core configurations looks exactly the same as for the Re and Fe alloys. On the contrary, alloying with Ta or Zr produces the same core symmetry as pure tungsten. The same result was obtained for W-Ta by Li {\it et al}~\cite{WRECORE_2}. The core structure of pure W is symmetric, as shown in Figure \ref{fig:CORE_STR}a, i.e., the dislocation expands equally along the six $\langle112\rangle$ directions, similar to results obtained earlier~\cite{WRECORE}.



Alloying with Re or any other TM with a higher number of valence electrons leads to a change in dislocation core structure from symmetric to asymmetric. The core is spread out in three $\langle112\rangle$ directions on $\{110\}$ planes \cite{Xu_Moriarty}. There are a six possible $\langle112\rangle$ orientations and {\color{black} the core is thus double degenerate}. The transition from symmetric to asymmetric core can change the dislocation slip plane~\cite{WRECORE}. The symmetric core dislocations glide uniformly on $\{110\}$ planes; asymmetric ones glide in a zigzag manner~\cite{Theory_of_disl} and the slip plane changes to $\{112\}$. Using the core site notation in Figure \ref{fig:CORE_STR}, {\color{black} the symmetric core glide path $1\rightarrow2\rightarrow4$, which is shown by the gray band in Figure \ref{fig:CORE_STR}a, will be changed to glide path $1\rightarrow2\rightarrow4\rightarrow3\rightarrow5$, shown by the gray band in Figure \ref{fig:CORE_STR}c~\cite{GROGER1} }. In the present work, the transition from symmetric to asymmetric core was obtained in the VCA calculations and this result was confirmed for the W-Re alloy in a super cell model calculation reported earlier~\cite{WRECORE, WRECORE_2}.

\begin{figure}
\includegraphics[angle=270, width=16.0cm]{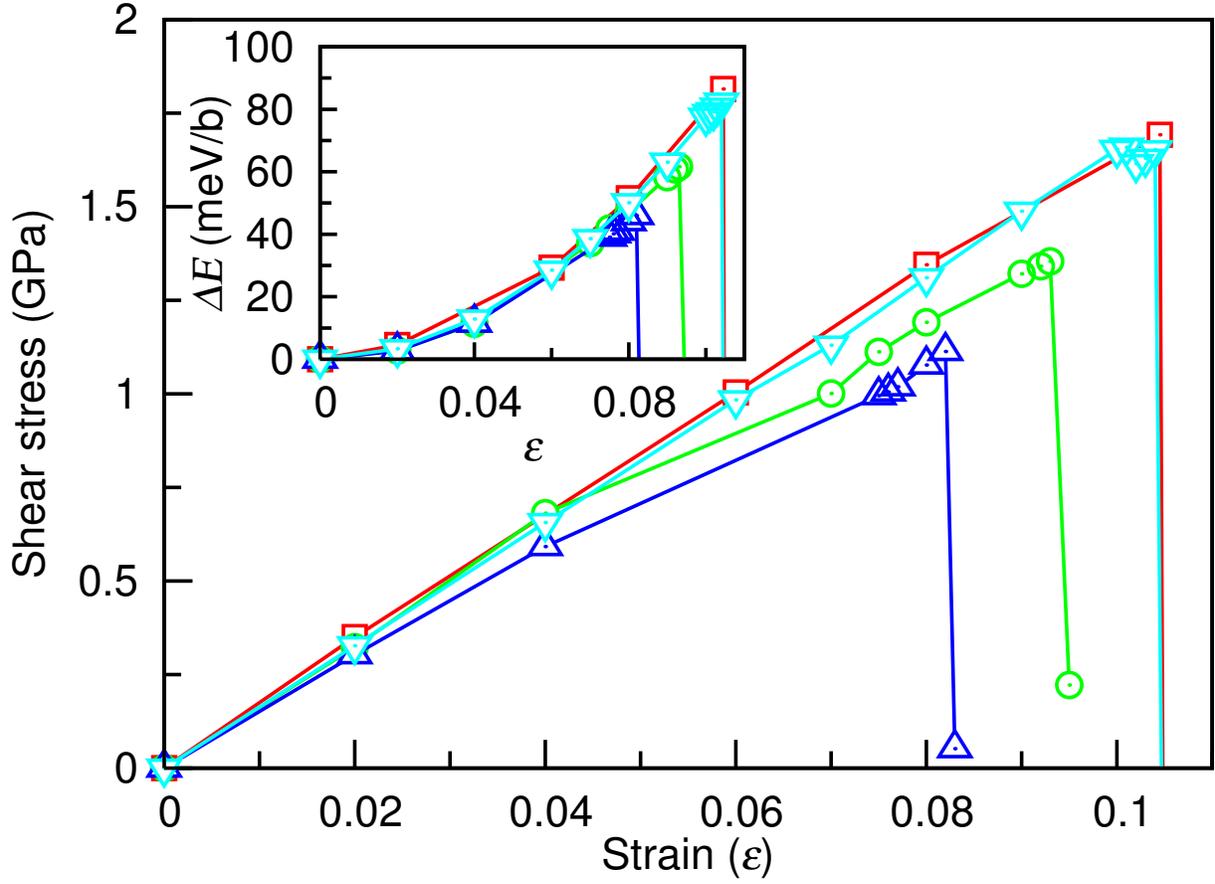}
\caption{Shear stress and total energy, $\Delta E$, per dislocation per Burgers vector $\vec b$ (shown in the inset) as a function of strain, $\varepsilon$, for pure $W$ (shown by squares), W$_{0.90}$Re$_{0.10}$ (shown by {\color{black} circles}), W$_{0.95}$Fe$_{0.05}$ (shown by up pointing triangles) and W$_{0.98}$Fe$_{0.02}$ (shown by down pointing triangles).}
\label{fig:PSTRESS}
\end{figure}

Addition of Re or any TM from group VIIIA leads to the change of dislocation slip plane and decrease in the value of Peierls stress $\sigma_p$~\cite{WRECORE}. Following a widely used technique~\cite{PEIERSL_STRESS}, {\color{black} we apply pure shear strain in the $\langle111\rangle$ direction which results in a stress that has $\sigma_{zx}$ as its main component, the influence of $\sigma_{yx}$ can be neglected since $\sigma_{yx}/\sigma_{zx}=0.04$ for pure W}. The corresponding strain is produced by modification of the basis vector  $\vec h_1$ along $\vec u_3$ direction $\vec h_1 = 1/2\vec u_1 + 1/2\vec u_2 + (1/2+\varepsilon)\vec u_3$. Figure~\ref{fig:PSTRESS} shows the dependence of total energy (see inset) per dislocation per Burgers vector and shear stress as a function of strain, $\varepsilon$. 



For small strains, i. e. in the elastic regime, the energy increases as $\varepsilon^2$. At larger $\varepsilon$ the energy dependence deviates from square dependence and abruptly drops. This drop in energy is caused by a jump of the dislocation core to {\color{black} the next stable "easy" core neighboring position ($1\rightarrow4$)}. The corresponding stress and strain are considered to be the Peierls stress $\sigma_p$ and strain $\varepsilon_p$. Earlier it was demonstrated~\cite{WRECORE} that the cell size of 135 atoms as used in the present work is enough to reach convergence in  $\sigma_p$ values. Alloying with both Re and Fe reduces values of $\varepsilon_p$ and $\sigma_p$. For Fe this reduction is even larger than for Re although the solute concentrations corresponds to the same $e/a$ value in both cases. The absolute value of $\sigma_p$ equal to 1.71 GPa for pure tungsten (Table \ref{PS})) is somewhat lower than those obtained earlier~\cite{WRECORE}, which were 2.09 GPa using QE and 2.37 GPa using VASP. {\color{black} The difference is attributed to the use of different types of pseudo-potentials, ultra-soft in our case and norm-conserving {\color{black} (QE) and PAW (VASP)} in Romaner {\it et al.}~\cite{WRECORE}, and the higher accuracy obtained using our larger plane wave energy cut-off of 42 Ry compared to 30 Ry {\color{black}(QE calculations) and 16.4 Ry (VASP calculations)}~\cite{WRECORE}. Moreover, the norm-conserving pseudopotential would require an even larger cut-off energy than the ultra-soft one.} The relative reduction of $\sigma_p$ is 20 \% for Re and 36 \% for Fe. A larger reduction of $\sigma_p$ in tungsten-iron alloys correlates with the larger reduction of $C^\prime$ modulus caused by alloying with Fe (Figure ~\ref{cp_a} and Table ~\ref{elst_cnst}). These results demonstrate that even if the reduction of Peierls stress and  $C^\prime$ values can be qualitatively understood as arising from a filling of $d$-states in tungsten by additional electrons from group VII and VIII solutes, some other element-specific mechanism exists. It should be mentioned that the solubility limit of Fe in W is around 2\% and in the concentration of Fe in limited to this value the reduction of Peierls barrier is about one half of that of the alloy with 5\% Fe. {\color{black} For the W-Ru alloy with a solubility limit of 3 \%, $\sigma_p$ is reduced by 6 \%. This result is consistent with the experimentally observed reduction of ductile-brittle transition temperature in W-Ru alloy~\cite{PLASMA_PROCESSES}.}

\begin{figure}

\includegraphics[angle=270,width=16.2cm]{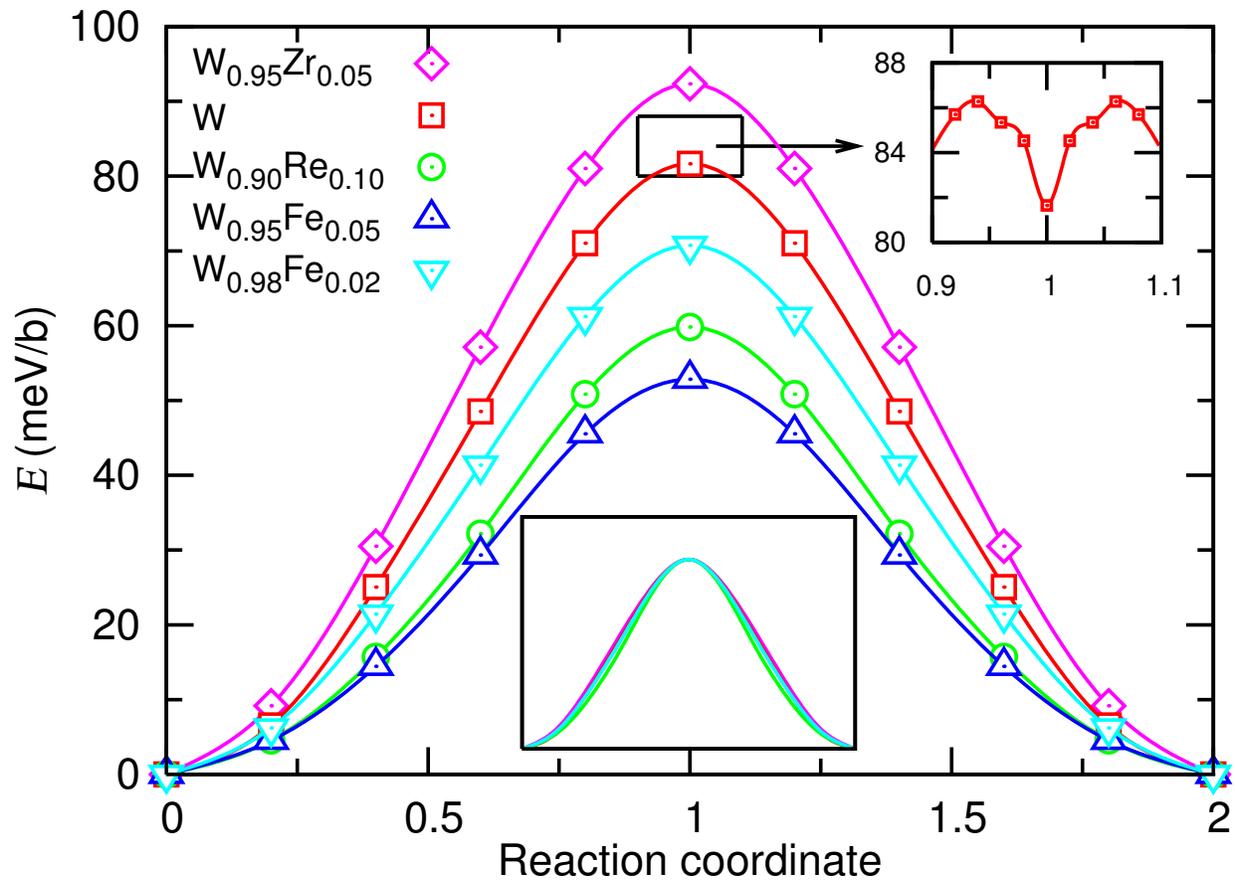}
\caption{Calculated Peierls barrier, where reaction coordinate, $r_c$, equal 0 for "easy" core configuration and 1 for "split" core. {\color{black}The curves are normalized in the center insert. The detailed structure of the calculated Peierls barrier in pure W around maximum is shown in the insert at upper right.}}
\label{fig:PBARRIER}
\end{figure}

\begin{figure}

\includegraphics[angle=0,width=14.0cm]{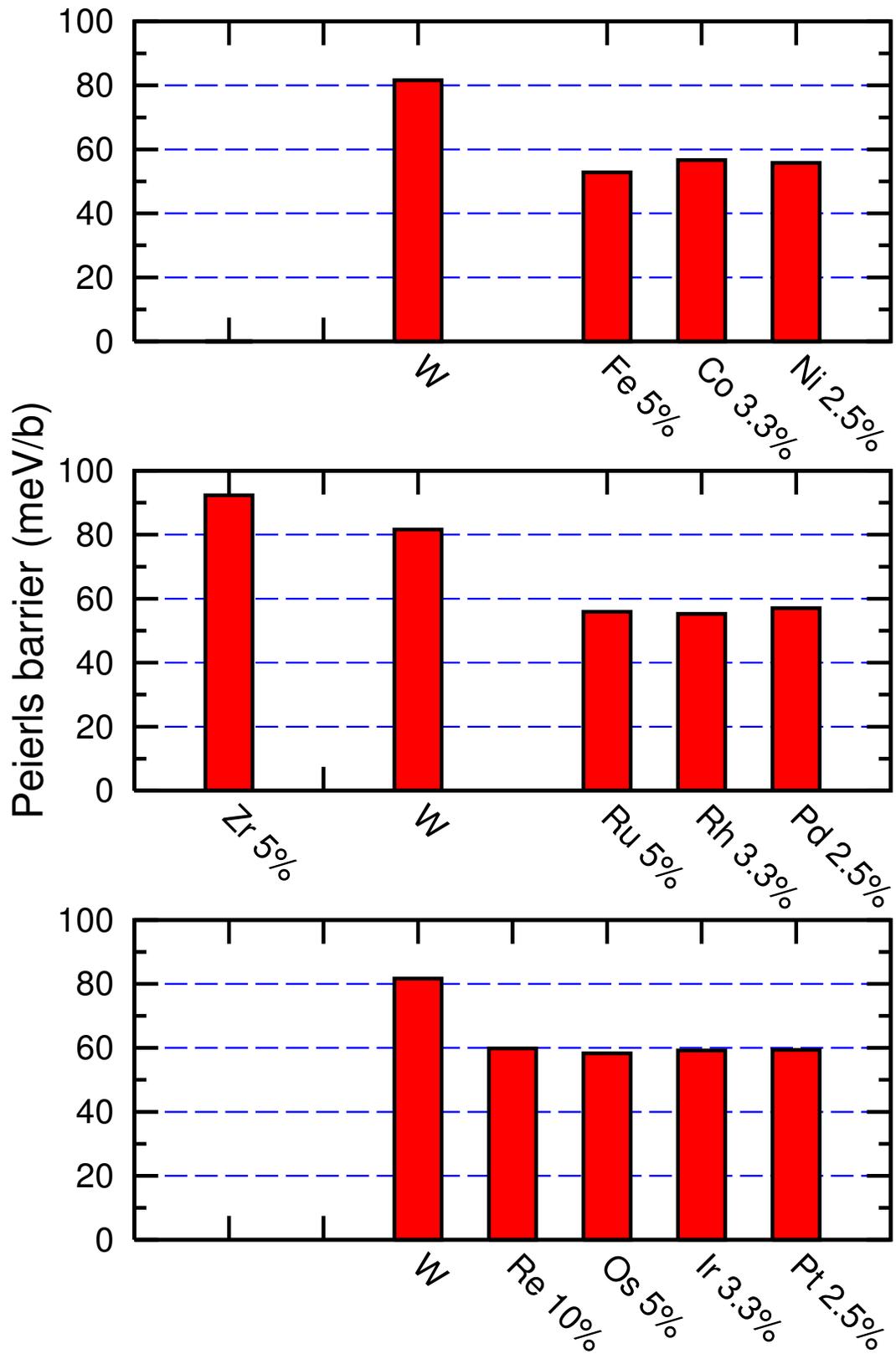}
\caption{The height of Peierls barrier for the set of W alloys.}
\label{fig:HPBARRIER}
\end{figure}

The second method used in present work to estimate Peierls stress was originally proposed by Nabarro~\cite{Nabarro}{\color{black}, and has been discussed in publications by Gr\"{o}ger {\it et al.}~\cite{GROGER1, GROGER2}}. Following Nabarro, we measure the system energy as a function of the dislocation location as it moves from position 1 to position 2 in Figure \ref{fig:CORE_STR}. {\color{black} It is easy to see that both $1\rightarrow2\rightarrow4$  ($\{110\}$ slip plane) and $1\rightarrow2\rightarrow4\rightarrow3\rightarrow5$ ($\{112\}$ slip plane) paths contain a set of dislocation core jumps between "easy" and "split" core configurations with the same $1\rightarrow2$ barrier.} The energy dependence obtained is associated with the Peierls barrier and the Peierls stress is given by the maximum gradient of this function
\begin{equation}
 \sigma_p = \frac{1}{b}\left(\frac{dV}{dR_c}\right)_{max},
\label{sigma_p}
\end{equation}
where $b$ is the Burgers vector, {\color{black} $V$ is energy per unit length of a straight dislocation and $R_c$ is the distance along the dislocation core path on the move from 1 to 2. $R_c$ equals 0 for position 1 and $\sqrt 2 a/3$, where $a$ is the lattice parameter, for the position 2. We use the drag method, also called  the reaction coordinate method~\cite{Xu_Moriarty}, to move the dislocation from 1 ("easy") to 2 ("split") core configuration. According to this method, the $z$-coordinates of atoms in two columns around both cores in the modeling cell, shown by purple circles in Fig.~\ref{fig:CORE_STR}a, are fixed so that $z=r_c z_0+(1-r_c)z_1$, where $z_0$ and $z_1$ correspond to the $z$-coordinate of the specified atoms in the "easy" and "split" core configurations respectively, and $r_c=R_c/(\sqrt 2 a/3$) is the so called reaction coordinate. {\color{black} In the initial implementation~\cite{Xu_Moriarty}, the authors fix the $z$-coordinate of atom which has the largest displacement resulting from the movement of the dislocation core. However from our experience the atomic relaxation convergence increases if we fix the $z$-coordinate of any two of the three atoms surrounding site 2 in Fig.~\ref{fig:CORE_STR}a. In order to avoid any changes in the elastic interaction between dislocations, both cores are moved simultaneously.} The results obtained for the Peierls barrier changes are presented in Figure \ref{fig:PBARRIER} for pure tungsten and alloys of tungsten with Re, Fe, Ru and Zr. In order to demonstrate that the "split" core configuration corresponds to a metastable state, we calculated the potential barrier using a small reaction coordinate step for pure tungsten around $r_c=1$. As can be seen from the results presented in the upper insert in Figure \ref{fig:PBARRIER}, the "split" core configuration corresponds to a local minimum in the potential energy.} When normalized using the values for pure tungsten, the shape of the Peierls barrier for the alloys is shown to be similar to tungsten and quite insensitive to solute type and concentration, see lower insert of Figure \ref{fig:PBARRIER}. In all these alloys, the results yield a single barrier. For alloys with Os, Ir, Pt, Ru, Rh, Pd, Co and Ni the total energy was calculated for the "easy" core (reaction coordinate, $r_c=0$) and "split" core  ($r_c=1$). Since the shape of Peierls barrier is insensitive to type of solute, the difference between these energies provides an estimate of the value of the Peierls barrier as shown in Figure \ref{fig:HPBARRIER}. The reduction of the Peierls barrier observed in these calculations correlates well with the reduction in Peierls stress estimated in the deformation simulation described above. Alloying with Re or any group VIII transition metal at a concentration which corresponds to $(e/a)=6.10$ reduces the barrier by ~25 \%. Fe reduces the Peierls barrier even more significantly, whereas Zr increases the barrier. \comments{A closely related reduction of stiffness for moving row in dislocation core with placed in it solute atoms with a higher number of valence $d$ electrons and increase of stiffness for the solute atoms with a lower number was obtained for Mo alloys with $5d$ transition metals} This is consistent with the effect obtained for Mo alloys with 5$d$ transition metals in which placing a row of solute atoms with higher number of valence $d$ electrons along the dislocation core reduce stiffness and a row of atoms with lower number increase stiffness~\cite{Trinkle}. 

{\color{black} The Peierls stress calculated using the direct deformation method, ${{\sigma}}_p$, and from the Peierls barrier using Eq.\ref{sigma_p}, ${{\tilde{\sigma}}}_p$ are presented in Table \ref{PS}. For all alloys the Peierls stress values obtained from the Peierls barrier are lower than those from the direct method. The difference is largest, 14 \%, for W alloys with 10 \% Re, or 2 \% Fe. Thus, the ${{\tilde{\sigma}}}_p$ results, introduction of  2 \% Fe to tungsten reduces the Peierls stress by 12 \%. While for the pure W and W alloy with 5 \% of Fe it is below 5 \%. Gr\"{o}ger and Vitek~\cite{GROGER2} connected the discrepancy in the results of two methods with the fact that both drag and Nudged Elastic Band~\cite{NEB} (NEB) methods garantee that images of the system are distributed uniformly along the minimum energy path and do not implay that dislocation position is distributed along this path. In order to overcome this problem, the authors introduced a modified NEB~\cite{GROGER2} method which gives a Peierls stress that agrees within 8 \% of the directly calculated stress. However we believe that the 14 \% accuracy obtained in our calculation is reasonably good considering that maintains the same trend in ${{\sigma}}_p$. It should be metioned that problems with direct application of drag or NEB methods can be more substentional. According to modeling result obtained in the deformed cell the dislocation core structure is slightly modified with increase of deformation and then jumps directly to next "easy" core position. The dislocation core is not observed in metastable, "split" core position. Thus the agreemnt between data presented in Table \ref{PS} gives a support to application of drag method to calculation of Peierls stress.} 

The reduction of Peierls stress/barrier due to alloying with transition metals with a higher number of valence electrons and the increase for lower number of valence electrons is supported by existing experimental data. Hardening was observed in W-Zr alloys~\cite{EXP_WZr} and softening in W-Re/Ir alloys~\cite{Luo}. Very interesting experimental results for Mo alloys with group VIIIA $3d$ transition metals were presented by Hiraoka {\it et al}~\cite{Hiraoka_Mo}. The authors report a decrease of yield strength with addition of any element from group VIIIA, Fe, Co, Ni or $4d$ Pd. The author concluded that this effect could be well understood in the terms number of valence electrons of alloying elements, and that the effect of atomic size mismatch is secondary and minor. Although the value of yield strength is not directly related to ductility, the observed correlation of properties with $e/a$ value supports the validity of our calculation using the VCA approach.


	


	

\section{Conclusions}
\label{sec:conclusions}

The influence of alloying tungsten with transition metal solutes on the elastic properties and properties of $1/2\langle111\rangle$ screw dislocation based on electronic structure calculations. In comparison with a standard supercell method, we demonstrated that the virtual crystal approximation gives a fairly good description of such alloying, especially in the case of 4$d$ and 5$d$ transition metals. For the case of alloying with transition metals from group VIIIA, the modification of elastic constants, Peierls stress and barrier can be understood within the rigid band approximation. This means that alloys with TM concentrations leading to the same number of electrons per atoms exhibit a similar reduction of $C^\prime$ modulus and elastic anisotropy $A$. Together with this modification of elastic moduli the value of the Peierls stress and barrier are reduced by alloying with transition metals from group VIIIA. In addition to the Peierls barrier reduction, alloying with any metal from this group changes the "easy" core dislocation structure from the symmetric to asymmetric configuration. This similarity allows us to conclude that the search for an alternative to Re could be reduced to transition metals from the group VIIIA.

\section{Acknowledgments}

Work was supported by the U.S. Department of Energy Office of Fusion Energy Sciences. This research used resources of the National Energy Research Scientific Computing Center, which is supported by the Office of Science of the U.S. Department of Energy.


\bibliography{tungsten}

\newpage

\listoffigures

\newpage

\listoftables

\newpage
\newpage


\newpage


\newpage


\newpage


\newpage


\newpage


\newpage


\newpage



\newpage

\renewcommand{\baselinestretch}{1.0}

\begin{table*} 
\caption{\color{black}Comparison of lattice parameter ($a$) in \AA, bulk modulus ($B$) and elastic constants ($C_{ij}$) in GPa calculated using the VCA and supercell (SC) approaches for W$_{1-x}$TM$_x$. The concentration $x=0.0625$ corresponds to W$_{15}$TM.}
\begin{tabular*}{0.70\textwidth}{@{\extracolsep{\fill}}lccccccc}
\hline 
TM &Approach& $a$ & $B$ & $C^\prime$ & $C_{44}$ & $C_{11}$ & $C_{12}$ \\  
\hline 
\hline
  pure W & VCA  & 3.1903 &  304  &  160  &  141  &  518  &  197 \\ 
      &  SC     & 3.1896 &  308  &  157  &  139  &  517  &  203 \\ 
\hline
  Zr  &  VCA    & 3.2095 &  287  &  153  &  126  &  491  &  185 \\ 
      &  SC     & 3.2079 &  286  &  150  &  127  &  489  &  186 \\ 
\hline
  Ta  &  VCA    & 3.1946 &  298  &  161  &  136  &  513  &  191 \\ 
      &  SC     & 3.1963 &  308  &  157  &  135  &  517  &  203 \\ 
\hline
  Fe  &  VCA    & 3.1802 &  284  &  132  &  128  &  459  &  196 \\ 
      &  SC     & 3.1716 &  300  &  140  &  132  &  487  &  207 \\ 
\hline
  Co  &  VCA    & 3.1788 &  284  &  125  &  115  &  450  &  200 \\ 
      &  SC     & 3.1725 &  297  &  129  &  131  &  469  &  211 \\ 
\hline
  Ni  &  VCA    & 3.1830 &  279  &  116  &  126  &  434  &  202 \\ 
      &  SC     & 3.1743 &  295  &  126  &  131  &  463  &  211 \\ 
\hline
  Ru  &  VCA    & 3.1772 &  308  &  142  &  142  &  498  &  213 \\ 
      &  SC     & 3.1842 &  301  &  135  &  137  &  481  &  211 \\ 
\hline
  Rh  &  VCA    & 3.1710 &  309  &  135  &  141  &  490  &  219 \\ 
      &  SC     & 3.1862 &  298  &  124  &  133  &  462  &  215 \\ 
\hline
  Pd  &  VCA    & 3.1639 &  310  &  129  &  139  &  482  &  224 \\ 
      &  SC     & 3.1901 &  293  &  116  &  131  &  448  &  216 \\ 
\hline
  Re  &  VCA    & 3.1857 &  308  &  151  &  144  &  510  &  207 \\ 
      &  SC     & 3.1859 &  308  &  150  &  143  &  508  &  208 \\ 
\hline
  Os  &  VCA    & 3.1809 &  312  &  143  &  146  &  503  &  216 \\ 
      &  SC     & 3.1848 &  307  &  136  &  144  &  489  &  216 \\ 
\hline
  Ir  &  VCA    & 3.1736 &  316  &  147  &  148  &  513  &  218 \\ 
      &  SC     & 3.1865 &  304  &  125  &  140  &  470  &  221 \\ 
\hline
  Pt  &  VCA    & 3.1687 &  317  &  132  &  148  &  494  &  229 \\ 
      &  SC     & 3.1917 &  299  &  113  &  135  &  449  &  223 \\ 
\hline

\label{elst_cnst_vca_vs_sc}
\end{tabular*} 
\end{table*}

\newpage

\renewcommand{\baselinestretch}{1.0}

\begin{table*} 
\caption{Experimental and calculated{\color{black}(based on VCA)} values of lattice parameter ($a$) in \AA, bulk modulus ($B$) and elastic constants ($C_{ij}$) in GPa for W$_{1-x}$TM$_x$. {\color{black} The concentration for all TM except Re is chosen to keep number of electrons per atom $(e/a)=6.10$}.}
\begin{tabular*}{0.70\textwidth}{@{\extracolsep{\fill}}lccccccc}
\hline 
TM          & $x$  & $a$ & $B$ & $C^\prime$ & $C_{44}$ & $C_{11}$ & $C_{12}$ \\  
\hline 
\multicolumn{8}{c}{experiment~\cite{C_IJ_WRE}} \\
\hline 
W           & 0.0000 & 3.1659 & 314 & 164 & 163 & 533 & 205 \\
\hline 
\multicolumn{8}{c}{calculated, virtual crystal approximation} \\
\hline
  Re          &  0.0000  & 3.1903 &  304  &  160  &  141  &  518  &  197 \\ 
              &  0.0300  & 3.1880 &  306  &  156  &  142  &  514  &  202 \\ 
              &  0.0500  & 3.1866 &  307  &  153  &  143  &  512  &  205 \\ 
              &  0.1000  & 3.1831 &  310  &  146  &  145  &  504  &  212 \\ 
              &  0.3000  & 3.1689 &  322  &  130  &  156  &  495  &  235 \\ 
              &  1.0000  & 3.1317 &  350  &  -30  &  163  &  309  &  370 \\ 
\hline
  Zr          &  0.0250  & 3.1972 &  297  &  161  &  136  &  511  &  190 \\ 
\hline
  Ta          &  0.0500  & 3.1936 &  299  &  162  &  137  &  515  &  191 \\ 
\hline
  Fe          &  0.0250  & 3.1849 &  296  &  148  &  136  &  494  &  198 \\ 
\hline
  Ru          &  0.0250  & 3.1846 &  306  &  153  &  141  &  510  &  204 \\ 
\hline
  Rh          &  0.0167  & 3.1846 &  306  &  153  &  141  &  509  &  204 \\ 
\hline
  Os          &  0.0250  & 3.1862 &  307  &  153  &  143  &  511  &  205 \\ 
\hline
  Ir          &  0.0167  & 3.1849 &  308  &  154  &  143  &  512  &  205 \\ 
\hline

\label{elst_cnst}
\end{tabular*} 
\end{table*}

\newpage

\renewcommand{\baselinestretch}{1.0}

\begin{table*} 
\caption{In order to compare the reduction of Peierls stress calculated using the direct deformation method, ${\sigma}_p$ in GPa, and from the Peierls barrier using Eq.\ref{sigma_p}, ${\tilde{\sigma}}_p$ in W, W$_{0.90}$Re$_{0.10}$, W$_{0.95}$Fe$_{0.05}$, W$_{0.98}$Fe$_{0.02}$ and W$_{0.97}$Ru$_{0.03}$. }
\begin{tabular*}{0.60\textwidth}{@{\extracolsep{\fill}}lccc}
\hline 
                      & $\sigma_p$ & ${\tilde{\sigma}}_p$ \\  
\hline 
W                     & 1.71       & 1.64                 \\
\hline 
W$_{0.95}$Zr$_{0.05}$ & 2.18       & 1.82                 \\
\hline
W$_{0.90}$Re$_{0.10}$ & 1.37       & 1.15                 \\
\hline 
W$_{0.95}$Fe$_{0.05}$ & 1.09       & 1.05                 \\
\hline 
W$_{0.98}$Fe$_{0.02}$ & 1.65       & 1.40                 \\
\hline 
W$_{0.97}$Ru$_{0.03}$ & 1.60       & 1.35                 \\
\hline 

\label{PS}
\end{tabular*} 
\end{table*}

\end{document}